\documentstyle[12pt]{article}
\newcommand{\be}{\begin{equation}}
\newcommand{\ee}{\end{equation}}
\newcommand{\bea}{\begin{eqnarray}}
\newcommand{\eea}{\end{eqnarray}}
\newcommand{\nen}{\nonumber \\ \relax}

\newcommand{\forcepar}{{\hskip 10pt\vskip -15pt}}
\newfont{\headfont}{cmbx10 scaled 1440}
\newfont{\headfontb}{cmbx10 scaled 1200}
\newfont{\namefont}{cmr9}
\newfont{\initialfont}{cmr9 scaled 1200}
\newfont{\addfont}{cmti10}

\newcommand{\seq}{\ =\ }
\newcommand{\pls}{\ +\ }
\newcommand{\mi}{\ -\ }
\newcommand{\seqv}{\ \equiv\ }
\newcommand{\half}{\frac{1}{2}}
\newcommand{\inv}[1]{\frac{1}{#1}}
\newcommand\mybox[2]{\vcenter{\hrule width#1in\hbox{\vrule height#2in
   \hskip#1in\vrule height#2in}\hrule width#1in}}
\newcommand\dal{\mybox{0.1}{0.1}}
\newcommand{\IR}{{I \kern -0.4em R}}
\newcommand{\IC}{{I \kern -0.65em C}}

\newcommand{\prd}[1]{{\it Phys. Rev.} {\bf D#1}}
\newcommand{\ap}[1]{{\it Ann. Phys.} {\bf #1}}
\newcommand{\np}[1]{{\it Nucl. Phys.} {\bf B#1}}
\newcommand{\cmp}[1]{{\it Commun. Math. Phys.} {\bf #1}}
\newcommand{\ijmp}[1]{{\it Intl. J. Mod. Phys.} {\bf A#1}}

\newcommand{\cqg}[1]{{\it Class. Quantum Grav.} {\bf #1}}
\newcommand{\mpl}[1]{{\it Mod. Phys. Lett.} {\bf A#1}}

\newcommand{\pr}[1]{{\it Phys. Rev.} {\bf #1}}
\newcommand{\pl}[1]{{\it Phys. Lett.} {\bf #1B}}

\newcommand{\susy}{{SU{\hskip -2pt}SY}}
\newcommand{\gmat}[3]{(\gamma^{#1})_{#2}{}^{#3}}
\newcommand{\ggmat}[1]{(\gamma#1)}

\newcommand{\cq}{{\cal Q}}
\newcommand{\cs}{{\cal S}}
\newcommand{\aaux}{\Lambda}
\newcommand{\baux}{\Sigma}
\newcommand{\ct}{{\cal T}_{\sl g}}

\newcommand{\gtwist}{{\bf{\sl g}}--twist }

\begin{document}
\begin{titlepage}
\renewcommand{\thefootnote}{\fnsymbol{footnote}}
\begin{center}
{\headfont Twisting to Abelian {\large\sl BF}/Chern-Simons
Theories}\footnote{This work is supported in part by funds provided by the
U. S. Department of Energy (D.O.E.) under contract
\#DE-AC02-76ER03069.}
\end{center}
\vskip 0.3truein
\begin{center}
{{\initialfont R}{\namefont OGER}
		    {\initialfont B}{\namefont ROOKS,}
{\initialfont J}{\namefont EAN-}{\initialfont G}{\namefont UY}
		    {\initialfont D}{\namefont EMERS AND}
{\initialfont C}{\namefont LAUDIO}
		    {\initialfont L}{\namefont UCCHESI}\footnote{Supported by the Swiss
National Science Foundation.}
}
\end{center}
\begin{center}
{\addfont{Center for Theoretical Physics,}}\\
{\addfont{Laboratory for Nuclear Science}}\\
{\addfont{and Department of Physics,}}\\
{\addfont{Massachusetts Institute of Technology}}\\
{\addfont{Cambridge, Massachusetts 02139 U.S.A.}}
\end{center}
\vskip 0.5truein
\begin{abstract}

Starting from  a $D=3$, $N=4$ supersymmetric theory for matter fields,  a
twist with a Grassmann parity change  is defined which maps the theory into a
gauge fixed, abelian  $BF$ theory on curved 3-manifolds.  After adding surface
terms to
this theory, the twist is seen to map the resulting supersymmetric action to
two  uncoupled copies of the gauge fixed Chern-Simons action.  In addition, we
give
a map which takes the $BF$ and Chern-Simons theories into  Donaldson-Witten
TQFT's.
A similar construction, but with $N=2$ supersymmetry, is given in two
dimensions.
\vskip 0.5truein
\leftline {CTP \# 2237  \hfill August 1993}
\smallskip
\leftline{hep-th/9308127}
\end{abstract}

\end{titlepage}
\setcounter{footnote}{0}
\section{Introduction}
\forcepar
This paper deals with the problem of mapping supersymmetric field theories
into topological field theories (TFT's) [1-6] and of mapping different classes
of TFT's among themselves.   TFT's fall under two classes.
The first of the TFT's are the Schwarz-type \cite{ASch},  commonly known as
$BF$, theories.   Chern-Simons theory in three dimensions is a special case of
$BF$ theory. The second
are Donaldson-Witten or Topological Quantum Field Theories (TQFT's) \cite{WIT}.
A sub-class of the TQFT's, the topological Yang-Mills (TYM)
theories are  gauge invariant.  Another sub-class of the TQFT's is given by the
topological sigma models which do not possess gauge invariances.

To date, these two classes of theories have had vastly different origins.  On
the one hand, the
$BF$ theories have  non-trivial classical actions and first order equations of
motion. Their classical (abelian) actions
on manifolds of dimension $D$ are metric independent as they are of the form
$\int_M B_{(k)}\wedge F_{(D-k)}$, where
$F_{(D-k)}=dA_{(D-k-1)}$ and the subscript denotes the form's degree.
These theories are  invariant under Maxwell (or Yang-Mills) gauge symmetries.
They are also symmetric under the $k$-form symmetry which shifts $B_{(k)}$ into
the exterior derivative of a $(k-1)$-form.
On the other hand, the TQFT's classical lagrangian are either $0$ or a
total derivative and are devoid of classical equations of motion.  Apart from
the possible surface term, the entire lagrangian of a TQFT is obtained
\cite{BS,BMS,LP} as a BRST gauge fixing of a symmetry (topological symmetry)
which manifests itself as compactly supported shifts of some field in the
theory (for example, the gauge field in TYM).  A large class of the latter
theories may also be obtained from $N=2$ \cite{WIT,BKE} or even $N\geq2$
\cite{Yam} supersymmetric theories via a procedure known as twisting.

We will work in three and two dimensions restricting ourselves to abelian BF
theories.  Placed in this context, we will
solve  a problem which has existed since the birth of these theories; namely,
how to obtain the BF theories via the twisting of some supersymmetric theory.
Furthermore, we will make substantial progress towards solving an equally
long-standing problem; namely, what (if any) is the relation between $BF$
theories and TQFT's.

As the twisting process will play an important role in our work, it is
appropriate to give a quick review \cite{WIT}  using the example of $\IR^4$.
Starting with
a $N=2$ supersymmetric field theory and writing the Lorentz group as
$SO_L(4)\simeq SU_l(2)\times SU_r(2)$, we then take the diagonal sum of
$SU_l(2)$ with the automorphism group of the $N=2$ superalgebra, $SU_I(2)$.
The result is a  $SU_d(2)$, which we use to form a new Lorentz group,
$SO_{L'}(4)\simeq SU_d(2)\times SU_r(2)$.  As a result, spin-$\half$ fields,
which also transformed as doublets of $SU_I(2)$, now become integer spinned,
Grassmann odd fields.  In three dimensions, the Lorentz group is $SO_L(3)\simeq
SU_L(2)$.  In oder to define a twist, the supersymmetric theory will have to
possess a $SU_I(2)$ automorphism group so that the new Lorentz group may be
taken to be the diagonal sum of the two $SU(2)$'s.  This means that the $D=3$
theory should be $N=4$ supersymmetric.  In two dimensions, we will require a
$U(1)$ automorphism group, hence an $N=2$ supersymmetric theory.

Glancing at the $BF$ lagrangian (see above), we see that the Grassmann even
fields are first order in derivatives.  Whereas, upon gauge fixing, the
Grassmann odd fields are second order in derivatives.  This is an inversion of
the usual structure in supersymmetric theories.  Scaling
this hurdle will be achieved by a second stage of the twisting wherein we will
change the Grassmann parity of the fields; (bosons) fermions will become
(anti-) commuting.  As the supersymmetric theory we will apply our twisting
procedure to  will not be gauge invariant,
the $BF$/Chern-Simons theory obtained will be gauge fixed.  In this way, we
will obtain the abelian $BF$ and Chern-Simons theories from $N=4$
supersymmetric theories in three dimensions. Similarly,  $N=2$ theories will be
twisted to  the abelian $D=2$ $BF$ theory.
As an artifact of the process, we will actually
obtain two (uncoupled) copies of Chern-Simons theory.

Previously, it had been
shown that the gauge fixed Chern-Simons theories \cite{DGS} (along with a
related construction for the $BF$  theories \cite{BM,GMS,Luc})  are invariant
under a
set of symmetries generated by a pair of scalar and a pair of vector
charges, all Grassmann odd.  The algebra of these charges allows a
$SL_I(2,\IR)\simeq SU_I(2)$ automorphism group.  The number of components of
these charges matches the number of components of four Majorana fermions and it
was  shown that this algebra  is a twisted version of a $D=3$, $N=4$
supersymmetry algebra\footnote{This algebra was termed $N=2$ in ref.
\cite{DGS}.}.  As
part of our work , we will find the missing $N=4$ supersymmetric theory which
realizes the untwisted algebra.   Since, as we will
show, the supersymmetric theory may also be twisted to a TQFT, we will then
formally relate  a subset of TQFT's to the abelian $BF$ theory.

Our paper is organized as follows.  In the next section, the two $N=4$
supersymmetric actions (which differ only by surface terms) we will use
throughout our three dimensional discussion will be presented.  Following this,
in
section \ref{TwistBF}, we will twist the first of these actions to the abelian
$BF$ theory in three dimensions.
After writing down the action for a $D=2$, $N=2$ scalar supermultiplet, we will
show how to twist this theory to the two dimensional $BF$ theory, in
sub-section
\ref{TwoDIM}.  In section \ref{TwistCS}, we shall return to three dimensions
and use the second action from
section \ref{ACTIONS},  which we will twist to the
abelian Chern-Simons theory.  The structure and
transformations generated by the three dimensional twisted superalgebra will be
given in section \ref{TwistSA}.
In section \ref{CTQFT}, we will show how to connect TQFT's
obtained from our supersymmetric theory with $BF$ theories via a change in
Grassmann parity.
We conclude in section \ref{CONC}.  The conventions used in this paper may be
found in the appendix.

\vskip 0.5truein\setcounter{equation}{0}
\section{The $N=4$ Supersymmetric Actions}\label{ACTIONS}
\forcepar
Let us begin by intoducing the two $N=4$ supersymmetric actions we will be
using in our discussion of the three dimensional topological theories.
In order to establish the main features of the twisting process it is best to
work on a flat manifold.  Later, we will extend the procedure to curved
manifolds (see sub-section (\ref{CURVE}).  Although the actions constructed in
this section exist in either
Minkowski  space-time or $\IR^3$, in the rest of the paper we will restrict our
discussion to manifolds with Euclidean signature.

Our  supersymmetric matter multiplet contains the following complex
fields:
\begin{center}
\begin{tabular}{||c|c|c||}\hline
{FIELD}	&SPIN&GRASSMANN PARITY\\ \hline\hline
$\phi$	&$0$&even\\ \hline
$\lambda$	&$0$&even\\ \hline
$\psi_\alpha$	&$1/2$&odd\\ \hline
$\chi_\alpha$	&$1/2$&odd\\ \hline
\end{tabular}
\end{center}
There are a number of possible actions we could write down for these fields.
Even within a given action, we can add surface terms.  We will see the
importance of this later.  As our basic action we take\footnote{The ordering of
the fields in the various terms is important since our twisting procedure
involves changing the Grassmann character of the fields.  We will take the
ordering as
given in this action throughout.}
\be
S^{\susy}\seq \int d^3x \big[ \partial^a {\bar \phi}\partial_a \lambda
\pls\partial^a \phi\partial_a {\bar \lambda}  \pls
i\half \chi^{\alpha }\gmat a \alpha\beta\partial_a{\bar \psi}_{\beta}
\mi i \half
{\bar \chi}^{\alpha }\gmat a \alpha\beta\partial_a{\psi}_{\beta}\big]
\ \ ,\label{EQNSF}
\ee
where the bar denotes complex conjugation.
This action is invariant under the following rigid supersymmetry
transformations\footnote{Throughout this paper we will discard  surface terms
while establishing the existence of  supersymmetries.}
\begin{center}
\begin{tabular}{rcllcl}
$[\cq_\alpha ,\phi] $&$\seq$&$  i\psi_\alpha\ \ ,$&
$[\cq_\alpha ,\lambda] $&$\seq$&$ i\chi_\alpha\ \ ,$\\
$\{\cq_\alpha ,{\bar\psi}_\beta\} $&$\seq$&$
-2\ggmat{^a}_{\alpha\beta}\partial_a
{\bar\phi}\ \ ,$&
$\{\cq_\alpha ,{\bar\chi}_\beta\} $&$\seq$&$
2\ggmat{^a}_{\alpha\beta}\partial_a
{\bar\lambda}\ \ ,$\\
$[\bar{\cq}_\alpha ,\bar\phi ] $&$\seq$&$  -i\bar{\psi}_\alpha\ \
,$&
$[\bar{\cq}_\alpha ,\bar \lambda] $&$\seq$&$  -i\bar{\chi}_\alpha\ \ ,$\\
$\{\bar{\cq}_\alpha ,{\psi}_\beta\} $&$\seq$&$
-2\ggmat{^a}_{\alpha\beta}\partial_a
{\phi}\ \ ,$&
$\{\bar{\cq}_\alpha ,{\chi}_\beta\} $&$\seq$&$
2\ggmat{^a}_{\alpha\beta}\partial_a
{\lambda}\ \ .$\\
\end{tabular}
\end{center}
\vskip -0.85truecm
\be
\label{PHANTOM}
\ee
The $\cq$-super--charges form the $N=2$ supersymmetry algebra
\be
\{ \bar{\cq}_\alpha,\cq_\beta\}\seq -i2 \ggmat{^a}_{\alpha\beta}\partial_a \ \
,\quad
\{ {\cq}_\alpha,\cq_\beta\}\seq 0\ \ ,\quad
\{ \bar{\cq}_\alpha,\bar{\cq}_\beta\}\seq 0\ \ .\label{EQNQALG}
\ee
The action is invariant under the interchange $\lambda\leftrightarrow\phi$.
{}From this, it follows that there is a  second $N=2$
supersymmetry of (\ref{EQNSF}),
\begin{center}
\begin{tabular}{rcllcl}
$[\cs_\alpha ,\phi] $&$\seq$&$  i\chi_\alpha\ \ ,$&
$[\cs_\alpha ,\lambda] $&$\seq$&$  i\psi_\alpha\ \ ,$\\
$\{\cs_\alpha ,{\bar\psi}_\beta\} $&$\seq$&$
-2\ggmat{^a}_{\alpha\beta}\partial_a
{\bar\lambda}\ \ ,$&
$\{\cs_\alpha ,{\bar\chi}_\beta\} $&$\seq$&$
2\ggmat{^a}_{\alpha\beta}\partial_a {\bar
\phi}\ \ ,$\\
$[\bar{\cs}_\alpha ,\bar\phi ] $&$\seq$&$  -i\bar{\chi}_\alpha\ \
,$&
$[\bar{\cs}_\alpha ,\bar\lambda] $&$\seq$&$  -i\bar{\psi}_\alpha\ \ ,$\\
$\{\bar{\cs}_\alpha ,{\psi}_\beta\} $&$\seq$&$
-2\ggmat{^a}_{\alpha\beta}\partial_a{\lambda}\ \ ,$&
$\{\bar{\cs}_\alpha ,{\chi}_\beta\} $&$\seq$&$
2\ggmat{^a}_{\alpha\beta}\partial_a
{\phi}\ \ ,$\\
\end{tabular}
\end{center}
\vskip -0.85truecm
\be
\label{EQNSTRANS}
\ee
The $\cs$-super--charges also form an $N=2$ supersymmetry algebra.
The automorphism group of each of the supersymmetry algebras is $U(1)$.

It will prove useful to re-write $S^\susy$ in terms of real/imaginary fermions
rather than the complex ones.  To do this, we define the real and imaginary
parts of the fermions via:
$\chi_\alpha\equiv \chi_{\alpha 1} + i \chi_{\alpha 2}$ and $\psi_\alpha\equiv
\psi_{\alpha 1} + i \psi_{\alpha 2}$.
Consequently, the action becomes
\be
S^\susy\seq \int d^3 x \big[\partial^a {\bar \phi}\partial_a \lambda
\pls\partial^a \phi\partial_a {\bar \lambda}\pls \chi^{\alpha
A}\gmat{a}{\alpha}{\beta}\partial_a\psi_\beta{}^B\epsilon_{AB}\big]\ \
.\label{EQNSFA}
\ee
The lagrangian in this action is equivalent to that in (\ref{EQNSF}); {\it
i.e.}, no surface terms were incurred in this re-writing.  In twisting to the
$BF$ theory, we will use this form of the action.

{}From $\psi_{\alpha A}$ and $\chi_{\alpha A}$, we can  construct another
action whose lagrangian differs from (\ref{EQNSF}) by a total derivative term.
To do this we define
$\Psi_{\alpha A}\equiv \psi_{\alpha A}+ i\chi_{\alpha A}$
($\bar\Psi_{\alpha}{}^{A}\equiv \psi_{\alpha}{}^{ A}- i\chi_{\alpha}{}^{ A}$).
As $\Psi_{\alpha A}$ is a complex doublet, we take it to transform as a {\bf 2}
of $SU_I(2)$ while $\bar\Psi_{\alpha}{}^{A}$ is in the conjugate
representation.
Using this in  the action (\ref{EQNSF})  we arrive at
\bea
S'{}^{\susy}&\seq& \int d^3x \big[\partial^a {\bar \phi}\partial_a \lambda
\pls\partial^a \phi\partial_a {\bar \lambda} \pls
i\half \bar\Psi^{\alpha B}\gmat a \alpha\beta\partial_a\Psi_{\beta B}\big]\nen
&\seq&S^{\susy}\pls ({\rm surface\  terms})
\ \ ,\label{EQNSFN}
\eea
and discard the surface terms.
The original two $N=2$ supersymmetries now become invariances of the action
under the following transformations
\bea
[Q_{\alpha A}, \phi]&\seq& i(\Psi_{\alpha A}\pls \bar\Psi_{\alpha A})\ \ ,\nen
[Q_{\alpha A},\lambda]&\seq& i(\Psi_{\alpha A}\mi \bar\Psi_{\alpha A})\ \ ,\nen
\{Q_{\alpha A},\Psi_{\beta B}\}&\seq& -2\epsilon_{AB} \ggmat{^a}_{\alpha\beta}
\partial_a(\bar\phi \mi \bar\lambda)\ \ ,\nen
\{Q_{\alpha A},\bar\Psi_{\beta B}\}&\seq& -2\epsilon_{AB}
\ggmat{^a}_{\alpha\beta}
\partial_a(\bar\phi\pls\bar\lambda)\ \ ,\nen
[{\bar Q}_{\alpha A}, \bar\phi]&\seq& -i(\bar\Psi_{\alpha A}\pls \Psi_{\alpha
A})\ \
,\nen
[{\bar Q}_{\alpha A},\bar\lambda]&\seq& -i(\bar\Psi_{\alpha A}\mi \Psi_{\alpha
A})\ \ ,\nen
\{{\bar Q}_{\alpha A},\bar\Psi_{\beta B}\}&\seq&-2\epsilon_{AB}
\ggmat{^a}_{\alpha\beta} \partial_a(\phi\mi\lambda)\ \ ,\nen
\{{\bar Q}_{\alpha A},\Psi_{\beta B}\}&\seq& -2\epsilon_{AB}
\ggmat{^a}_{\alpha\beta}
\partial_a(\phi\pls\lambda)\ \ .\label{EQNQA}
\eea
This shows explicitly that the both actions, $S^{\susy}$ and $S'{}^{\susy}$ are
invariant under  an $N=4$ supersymmetry.
 Indeed, the algebra of charges defined by (\ref{EQNQA}) is
\be
\{{\bar Q}_{\alpha}{}^{ A},Q_{\beta B}\}\seq i4 \delta_{B}{}^A
\ggmat{^a}_{\alpha\beta}
\partial_a\ \ .\label{EQNSALGF}
\ee
This algebra has a $SU_I(2)$ automorphism invariance with the $Q_{\alpha A}$
transforming

in the doublet representation.

\vskip 0.5truein\setcounter{equation}{0}
\section{Mapping to $BF$ Theories}\label{TwistBF}
\forcepar
This section is divided into three parts.  First, in sub-section
(\ref{TSSUSY}), we present the twisting
procedure while working with the action $S^\susy$.  As advertised, we will find
the twisted
action to be the gauge fixed, abelian $BF$ theory on $\IR^3$.  Then,  in
sub-section (\ref{CURVE}), we will
discuss how to obtain the $BF$ theory on curved manifolds.  Finally, in
sub-section (\ref{TwoDIM}), as another
example of the
procedure, we will write down a $D=2$, $N=2$ supersymmetric action from which
the two-dimensional abelian $BF$ theory may be obtained via twisting.
\goodbreak\subsection{{\bf{\sl g}}--Twisting $S^{\susy}$}\label{TSSUSY}
\forcepar

The Lorentz algebra in three dimensions is $SO_L(3)\simeq SU_L(2)$.  As the
first stage of our twisting we take all internal indices to be $SU_L(2)$
indices.  This amounts \cite{DGS} to re-defining the Lorentz group  to be the
diagonal subgroup of $SU_L(2)\times SU_I(2)$.  With this, the original scalar
fields remain Lorentz singlets while the real spin-$\half$ fields become
Lorentz bi-spinors: $\psi_{\alpha B}\to \psi_{\alpha\beta}$ and $\chi_{\alpha
B}\to \chi_{\alpha \beta}$.  This means that we can decompose $\psi_{\alpha
\beta}$ as a real vector plus a scalar field; similarly for $\chi_{\alpha
\beta}$.

As the second stage of our twist, we declare the fields to have
opposite Grassmann parity to those of the parent  supersymmetric theory.
This second step does not exist
in the known \cite{WIT} twisting of supersymmetric theories to obtain
Donaldson-like topological quantum field theories (TQFT's).
We call this two stage mapping a ``\gtwist'' and define it by the map
\bea
\ct&:&\  \psi_{\alpha B}\rightarrow \ \psi_{\alpha \beta}\seqv
\inv{\sqrt{2}}\big[ i\ggmat{^a}_{\alpha\beta} A_a \mi C_{\alpha\beta}
\baux\big]\
\ ,\nen
\ct&:&\  \chi^{\alpha B}\  \rightarrow \ \chi^{\alpha \beta}\seqv
\inv{\sqrt{2}}\big[ \ggmat{^a}^{\alpha\beta} B_a \pls iC^{\alpha\beta}
\aaux\big]\
\ ,\nen
\ct&:&\ \phi \ \rightarrow\   \inv{\sqrt{2}}(c \mi i b')\ \ ,\nen
\ct&:&\ \bar\phi \ \rightarrow\ \inv{\sqrt{2}}( c\pls i b')\ \ ,\nen
\ct&:&\ \lambda \ \rightarrow\ \inv{\sqrt{2}}(c'\pls i b)\ \ ,\nen
\ct&:&\ \bar\lambda \ \rightarrow\ \inv{\sqrt{2}}(c'\mi i b)\ \ ,\nen
\ct&:&\ \ \epsilon_{AB}\ \rightarrow\ iC_{\alpha\beta}
\ \ .\label{EQNTWIST}
\eea
The fields on the right hand side of the arrows are defined by this map to have
Grassmann parity opposite to those on the left.  The factors of ``$i$'' have
been inserted so that the process of complex
conjugation commutes with $\ct$.  Additionally, the other numerical factors are
for later convenience.  We summarize the new field
content in the following table:
\begin{center}
\begin{tabular}{||c|c|c||}\hline
{FIELD}	&SPIN&GRASSMANN PARITY\\ \hline\hline
$A_a$	&$1$&even\\ \hline
$\baux$	&$0$&even\\ \hline
$B_a$	&$1$&even\\ \hline
$\aaux$	&$0$&even\\ \hline
$c$	&$0$&odd\\ \hline
$b$	&$0$&odd\\ \hline
$c'$	&$0$&odd\\ \hline
$b'$	&$0$&odd\\ \hline
\end{tabular}
\end{center}

Performing the
map, $\ct$,  on the action $S^{\susy}$ as given in eqn. (\ref{EQNSFA})  we
find, up to
surface terms,
\be
S_{BF}\seq \int d^3 x \big[ \epsilon^{abc} B_a \partial_b A_c \pls
(\partial^aA_a)\aaux\pls (\partial^a B_a) \baux\pls   c'\dal c \pls  b'\dal
b\big]\
\ .\label{EQNSFT}
\ee
This is the action of  the fully gauge fixed abelian $BF$ theory in
three dimen--sions\footnote{The ordering of the fields in the gauge fixing
terms
is chosen so as not to introduce additional minus signs when we later map
to the TQFT.}.  The first term is the classical  $BF$ action.  In this term,
the Levi-Cevita tensor arises from a trace on the product of three gamma
matrices.  The second
and third terms represent the gauge fixings of the local $U(1)$ and $1$-form
symmetry on $B_a$ (see section (\ref{TwistSA}) for details).  In these terms,
the Lorentz dot product arises from the
trace of products of two gamma matrices.
The ghost actions for these gauge fixings are given by the
last two terms in (\ref{EQNSFT}).  Note that only the Landau gauge appears in
this procedure.  The surface terms mentioned
above appear only from the gauge fixing and ghost terms.  They are needed in
order to write these terms in their conventional forms.

\newpage
\subsection{Curved 3-Manifolds}\label{CURVE}
\forcepar
The classical $BF$ action is topological.  It is only after gauge fixing that a
metric appears in the action.  We would like to recover this peculiar metric
dependence.

We could simply \gtwist the action $S^\susy$ on $\IR^3$ to obtain
(\ref{EQNSFT}) and then covariantize it with respect to some background metric
on a curved manifold, $M$.  By definition, the subsequent action,
\be
S_{BF}^M\seq \int d^3 x \epsilon^{abc} B_a \partial_b A_c \pls \int d^3x
\sqrt{g}\big[
(\nabla^aA_a)\aaux \pls (\nabla^a B_a)\baux \pls   c'\triangle c \pls
b'\triangle
b\big]\
\ ,\label{EQNBFT}
\ee
is the gauge fixed $BF$ theory on $M$. The derivative $\nabla_a$ is covariant
with respect to diffeomorphisms of $M$: $\nabla_a \equiv e_a{}^m\partial_m +
\omega_a{}^bJ_b$.  Here
$e_a{}^m$ is the driebein with determinant $e$.  The object $\omega_a{}^b(e)$
is
the dual of the Lorentz spin-connection for which the dual of the Lorentz
generator is  $J_a$.

Instead, suppose we started with the $N=4$ gauged
supergravity\footnote{The construction of $D=3$, $N=4$ gauged
supergravity along with its explicit couplings to matter is beyond the scope of
this work.} version of  $S^\susy$.  Among the new fields introduced would be
four gravitini and a $SU_I(2)$ gauge field, $V_a$.  As an example, the
gravitini appear in the spin-connection in the covariant derivative.  The
latter is also covariant with respect to local $SU_I(2)$ gauge transformations
due to the introduction of $V_a$.  The action (\ref{EQNBFT}) does not contain
either of these fields as it is neither $N=4$ locally supersymmetric or
$SU_I(2)$ gauge invariant.   Thus, in the {{\bf{\sl g}}}--twisting, we must set
the gravitini to zero.
In order to maintain this ansatz, however, we must restrict the local
supersymmetry of the action so that the gravitini may not be transformed away
from zero.  Since the local supersymmetry variations of the gravitini,
$\zeta_{a\alpha}{}^A$, are given by the covariant derivative of the local
supersymmetry parameter,
we must find a covariantly constant anti-commuting parameter:
\be
\delta \zeta_{a\alpha}{}^A\seq D_a\epsilon_\alpha^A\seq \partial_a
\epsilon_\alpha^A\mi  \omega_{a \alpha}{}^\beta \epsilon_\beta{}^A \pls
V_{a B}{}^A\epsilon_\alpha^B\seq 0\ \ .
\ee
To do this, we accentuate our procedure in analogy with the twisting in $D=4$,
$N=2$ conformal supergravity backgrounds \cite{KR}.  We introduce a scalar
anti-commuting parameter, $\epsilon$ by $\epsilon_{\alpha}{}^A\equiv \epsilon
\delta_\alpha{}^A$ having  embedded the $SU_I(2)$ gauge field in the $SU(2)$
spin connection:   $\omega_{a\alpha}{}^\beta\delta_\beta{}^A\equiv V_{a
B}{}^A\delta_\alpha{}^B$.   All  supersymmetries are then lost with the
exception of the one generated by the scalar charges.  The corresponding
transformations  will be given later.  We then identify this curved background
with the geometry of $M$.

\goodbreak\subsection{Two Dimensions}\label{TwoDIM}
\forcepar

To illustrate the generality of our {{\bf{\sl g}}}--twisting procedure, we
offer an example in
two dimensions.
As the Lorentz group in two dimensions is $U(1)$, our supersymmetric theory
must have this abelian automorphism group.  This means that the theory must be
$N=2$ supersymmetric.  As our action  we take
\be
S^{\susy}_{D=2}\seq \int d^2 x \big[\partial^a\phi \partial_a\lambda \pls
i\half
\bar\psi^\alpha \ggmat{^a}_{\alpha}{}^{\beta}\partial_a \psi_\beta\big]\ \
,\label{EQNS2D}
\ee
where $\phi$ and $\lambda$ are  scalar fields and $\psi$ is a complex
spin-$\half$ field.  This action is invariant under the supersymmetry
transformations,
\bea
[Q_\alpha,\phi] &\seq& i \psi_\alpha\ \ ,\qquad\qquad\qquad\ \  [\bar
Q_\alpha,\lambda] \seq i
\bar\psi_\alpha\ \ ,\nen
[Q_\alpha,\bar\psi_\beta] &\seq& -2\ggmat{^a}_{\alpha\beta} \partial_a\phi\ \
,\qquad [\bar Q_\alpha,\psi_\beta]
\seq-2\ggmat{^a}_{\alpha\beta}\partial_a\lambda\ \ .\label{EQN2DSUS}
\eea
These form the $D=2$, $N=2$ supersymmetry algebra
\be
\{\bar Q_\alpha,Q_\beta\}\seq -i2\ggmat{^a}_{\alpha\beta}\partial_a\ \
.\label{EQN2DALG}
\ee

Upon defining $\psi_\alpha= \psi_{\alpha 1} + i\psi_{\alpha 2}$  and denoting
the new fermions as $\psi_{\alpha A}$,  $A=1,2$, we define the \gtwist to be
 \bea
\ct&:&\  \psi_{\alpha B}\rightarrow \ \psi_{\alpha \beta}\seqv
\big[ i\ggmat{^a}_{\alpha\beta} A_a \mi \ggmat{^3}_{\alpha\beta} B \mi \half
C_{\alpha\beta} \aaux\big]\
\ ,\nen
\ct&:&\ \phi \ \rightarrow\   c\ \ ,\nen
\ct&:&\ \lambda \ \rightarrow\ c'
\ \ .\label{EQN2DTWIST}
\eea
The $D=2$ analog of the procedure discussed in the previous sub-section but
with
$SU(2)$ replaced by $U(1)$,  may now be applied to the action (\ref{EQNS2D}).
It results in  the {\bf{\sl g}}--twisted action
\be
S_{BF}^{D=2}\seq 2\int d^2x \epsilon^{ab} B \partial_aA_b\pls
\int d^2x \sqrt{g} \big[ (\nabla^a A_a)\aaux\pls c'\triangle c\big]\ \
.\label{EQN2DBF}
\ee
This is the gauged fixed, abelian $BF$ action in two dimensions.

\vskip 0.5truein\setcounter{equation}{0}\goodbreak
\section{Mapping to Chern-Simons}\label{TwistCS}
\forcepar
As is well known, Chern-Simons theory is a special case of a $BF$ theory in
which the $A_a$ and $B_a$ fields are identified\footnote{In order to get the
non-abelian Chern-Simons theory, a term which is cubic in the $B_a$ field must
be added to the $BF$ lagrangian.}.  At the level of the fields
this is a purely formal operation.  However, when one considers that $A_a$ is a
$U(1)$-valued gauge field and $B_a$ is a singlet under that gauge group, one
realizes the absence of a representation theory prescription for the
identification.  At the level of symmetries, both fields transform as the
exterior derivative of a scalar parameter.  Thus, Chern-Simons theory is
strictly a special case of $BF$ theory only at the level of the structure of
the fields in the action.  Although this is a phenomenon in the gauge sector of
the theory, we might expect  similar behaviour with the space-time symmetries,
if we try to obtain
Chern-Simons via the {{\bf{\sl g}}}--twisting of a supersymmetric theory.
Indeed, we will see
that if we use the naive version of $S^{\susy}$, there is no group theoretic
prescription, in terms of $SU(2)$
representations, for the \gtwist.  Our map will be purely in terms of the
fields.
After seeing this, we will then turn to $S'{}^{\susy}$ (in the second
sub-section), for which both the
twist on the fields and the group theoretic interpretation are available.

\goodbreak\subsection{{\bf{\sl g}}--Twisted $S^{\susy}$ with $\chi$ and $\psi$
Identified}
\forcepar
Since we already know that $A_a$ and $B_a$ must be identified, we start by
identifying $\chi$ and $\psi$ in eqn. (\ref{EQNSFA}) so that we take the action
to be
\be
S^{\susy}_{0}\seq  \int d^3 x \big[ \partial^a\phi\partial_a\lambda \pls
\half\psi^{\alpha B}\gmat{a}{\alpha}{\beta}\partial_a\psi_{\beta B}\big]\ \
.\label{EQNSCS}
\ee
Here, $\lambda$ and $\phi$ are now real bosons\footnote{The real parts of the
corresponding fields from the previous sub-sections.} and $\psi^{\alpha A}$
represents a pair of real  spin$-\half$ fields, $A=1,2$.
Naively, we might define the \gtwist by the first line in eqn. (\ref{EQNTWIST})
along with
$\ct:\lambda\rightarrow c'$ and
$\ct:\phi\rightarrow c$.  Using this in $S^{\susy}_{CS}$ and applying the
procedure outlined in sub-section (\ref{CURVE}),
we arrive at the action
\be
S_{CS}\seq \half \int d^3 x \epsilon^{abc} A_a \partial_b A_c \pls \int
d^3x
\sqrt{g}\big[
(\nabla^aA_a)\baux\pls   c'\triangle c \big]\
\ .\label{EQNSCT}
\ee
Once again, we have switched the Grassmann parity of the fields.
Of course, this is the gauge fixed abelian Chern-Simons action.

As there are only two real fermions in this action, there is only a global
$SO(2)$ invariance, not $SU(2)$.  Thus we are unable to associate the Lorentz
symmetry of $S_{CS}$ with the diagonal sum of two $SU(2)$'s and there is
no group theoretic justification for taking the internal index on the fermions
to be Lorentz spinor indices, in the definition of the twist.  However,  we
simply point out that if this is done at the level of the fields, then the
Chern-Simons action is obtained.

\goodbreak\subsection{{\bf{\sl g}}--Twisting $S'{}^{\susy}$}
\forcepar
There is, however, a way to obtain the Chern-Simons action -- actually two
copies -- while having a group theoretic justification.  We start with
the action $S'{}^{\susy}$ (\ref{EQNSFN}) which differs from $S^{\susy}$ by
surface terms.  Now we take the internal $SU_I(2)$ indices on $\Psi_{\alpha A}$
to be Lorentz spin-$\half$ indices.  Again this amounts to re-defining the
Lorentz group to be the diagonal sum of the two $SU(2)$'s.  Then the \gtwist is
defined by
\be
\ct:\Psi_{\alpha B}\ \rightarrow\ \Psi_{\alpha \beta} \equiv \inv{\sqrt{2}}
\big[\ggmat{^a}_{\alpha\beta} (A_a+ iB_a) \pls iC_{\alpha\beta}
(\baux+i\aaux)\big]\ \
,\label{EQNTWCS}
\ee
along with a change of Grassmann parity.  $\ct$ acts on the scalar fields as
before (\ref{EQNTWIST}).  Performing these replacements in $S'{}^{\susy}$ and
applying the procedure outlined in sub-section (\ref{CURVE}), we
obtain
\bea
S^2_{CS}\seq&-\half\int d^3 x&\big[\epsilon^{abc} A_a\partial_b A_c\pls
\epsilon^{abc}B_a\partial_b B_c\big]\nen
&-\int d^3 x \sqrt{g}& \big[(\nabla^aA_a)\baux\pls  (\nabla^aB_a)\aaux\mi
c'\triangle
c\mi b'\triangle b\big]\ \ ,\label{EQNSCSA}
\eea
This is the action for two uncoupled copies of the gauge fixed Chern-Simons
theory.  Curiously, the appearance of more than one  gauge field is a phenomena
in extended supersymmetric Chern-Simons theories \cite{NG}.
Identifying the set of fields $(B_a,\aaux,b',b)$ with the set
$(A_a,\baux,c',c)$
reduces this to (twice) the action for one Chern-Simons gauge field
(\ref{EQNSCT}).

\newpage
\section{The {\bf{\sl g}}--Twisted Super-Algebra}\label{TwistSA}
\forcepar

In the context of gauge fixed theories, ``supersymmetry'' is to be understood
as a set of transformations generated by Grassmann odd charges which take
fields of ghost number $n$ into fields of ghost number $n\pm1$.
Vector super--charges of ghost number $1$ were  discovered for the
three--dimensional Chern-Simons theory
in the Landau gauge in ref. \cite{BRT}.  It was soon thereafter realized that
the same theory is further invariant under the anti-BRST transformations and
another vector generator both of ghost number $-1$ \cite{DGS}.  The BRST
generator and the ghost number $-1$ vector generator were found to close on
translations, thereby forming an $N=2$ supersymmetry algebra.  In addition, the
anti-BRST generator and the ghost number $1$ generator form another $N=2$
superalgebra.   The $N=2$ algebra, including the $BRST$ generator, was then
found to hold for the two-- and four--dimensional non-abelian $BF$ theories
\cite{BM,GMS}, and was generalized to arbitrary dimensions in ref. \cite{Luc}.
It was used to prove  the perturbative finiteness of the $D=3$
Chern-Simons theory \cite{DLPS} and of the $BF$ theory (see \cite{Luc} and
references
therein).
We will now extract these charges and algebras from our $N=4$ supersymmetry
algebra (\ref{EQNSALGF}) via twisting.

The \gtwist acts on the super--charges as
\bea
\ct &:&\  Q_{\alpha B}\ \rightarrow \  Q_{\alpha \beta}\seqv
\ggmat{^a}_{\alpha\beta} Q_a \pls iC_{\alpha\beta} Q\ \ ,\nen
\ct &:&\  {\bar Q}_{\alpha B}\ \rightarrow \  {\bar Q}_{\alpha \beta}\seqv
\ggmat{^a}_{\alpha\beta} {\bar Q}_a \pls iC_{\alpha\beta} {\bar Q}\ \
.\label{EQNQT}
\eea
In the absence of covariantly constant vectors, only the scalar super--charges
are conserved on curved manifolds.   On $\IR^3$, the full set of
super--charges is conserved.
Note that since the supercurrents were originally a product of a Grassmann
odd and the derivative of a Grassmann even field, the Grassmann parity of the
super-charges remains the same, namely odd.

Performing the map on the $N=4$ supersymmetry algebra (\ref{EQNSALGF}) we find
the {{\bf{\sl g}}}--twisted algebra whose  only non-trivial anti-commutators
are
\be
\{{\bar Q}_a,Q_b\}\seq -i2\epsilon_{abc}\partial^c\ \ ,\qquad \{{\bar
Q}_a,Q\}\seq
-i2\partial_a\ \ ,\qquad \{\bar Q,Q_a\}\seq i2\partial_a\ \ .\label{EQNTALG}
\ee
$Q$ and its complex conjugate are nilpotent.

The supersymmetry transformations (\ref{EQNQA})  now take the forms:
\begin{center}
\begin{tabular}{rcllcl}
$[Q,A_a]$&$\seq$&$ \partial_a(c+ib')\ \ ,$&
$[Q,B_a]$&$\seq$&$ i\partial_a(c'-ib)\ \ ,$\\
$[Q,\aaux]$&$\seq$&$0\ \ ,$&
$[Q,\baux]$&$\seq$&$0\ \ ,$\\
$\{Q,c\}$&$\seq$&$ i\baux\ \ ,$&
$\{Q,b\}$&$\seq$&$ i\aaux\ \ ,$\\
$\{Q,c'\}$&$\seq$&$ -\aaux\ \ ,$&
$\{Q,b'\}$&$\seq$&$ -\baux\ \ ,$\\
$$&$$& $$\\
$[\bar Q,A_a]$&$\seq$&$ \partial_a(c-ib')\ \ ,$&
$[\bar Q,B_a]$&$\seq$&$ -i\partial_a(c'+ib)\ \ ,$\\
$[\bar Q,\aaux]$&$\seq$&$0\ \ ,$&
$[\bar Q,\baux]$&$\seq$&$0\ \ ,$\\
$\{\bar Q,c\}$&$\seq$&$ -i\baux\ \ ,$&
$\{\bar Q,b\}$&$\seq$&$ -i\aaux\ \ ,$\\
$\{\bar Q,c'\}$&$\seq$&$ -\aaux\ \ ,$&
$\{\bar Q,b'\}$&$\seq$&$ -\baux\ \ ,$\\
$$&$$& $$\\
$[Q_a,A_b]$&$\seq$&$ -\epsilon_{abc}\partial^c(c+ib')\ \ ,$&
$[Q_a,B_b]$&$\seq$&$ -i\epsilon_{abc}\partial^c(c'-ib)\ \ ,$\\
$[Q_a,\aaux$&$\seq$&$ -i\partial_a(c'-ib)\ \ ,$&
$[Q_a,\baux]$&$\seq$&$ -\partial_a(c+ib')\ \ ,$\\
$\{Q_a,c\}$&$\seq$&$ iA_a\ \ ,$&
$\{Q_a,b\}$&$\seq$&$ iB_a\ \ ,$\\
$\{Q_a,c'\}$&$\seq$&$ -B_a\ \ ,$&
$\{Q_a,b'\}$&$\seq$&$ -A_a\ \ ,$\\
$$&$$& $$\\
$[\bar Q_a,A_b]$&$\seq$&$ -\epsilon_{abc}\partial^c(c-ib')\ \ ,$&
$[\bar Q_a,B_b]$&$\seq$&$ i\epsilon_{abc}\partial^c(c'+ib)\ \ ,$\\
$[\bar Q_a,\aaux]$&$\seq$&$ i\partial_a(c'+ib)\ \ ,$&
$[\bar Q_a,\baux]$&$\seq$&$ -\partial_a(c-ib')\ \ ,$\\
$\{\bar Q_a,c\}$&$\seq$&$ -iA_a\ \ ,$&
$\{\bar Q_a,b\}$&$\seq$&$ -iB_a\ \ ,$\\
$\{\bar Q_a,c'\}$&$\seq$&$ -B_a\ \ ,$&
$\{\bar Q_a,b'\}$&$\seq$&$ -A_a\ \ .$\\
\end{tabular}
\end{center}
\vskip -0.9truecm
\be\label{EQNTRANS}
\ee
These are symmetry transformations for the three--dimensional,  gauge fixed
$BF$ action.
Upon defining $Q\equiv s+ is'$ we find the BRST ($s$) and anti-BRST ($ s'$)
transformations to be
\begin{center}
\begin{tabular}{rcllcl}
$[s,A_a]$&$\seq$& $\partial_a c\ \ ,$& $[s,B_a]$&$\seq$&$ \partial_a b\ \ ,$\\
$[s,\aaux]$&$\seq$&$0\ \ ,$& $[s,\baux]$&$\seq$&$ 0\ \ ,$\\
$\{s,c\}$&$\seq$&$0\ \ ,$& $\{s,b\}$&$\seq$&$ 0\ \ ,$\\
$\{s,c'\}$&$\seq$&$ -\aaux\ \ ,$& $\{s,b'\}$&$\seq$&$-\baux\ \ ,$\\
$$&$$\\
$[s',A_a]$&$\seq$&$ \partial_a b'\ \ ,$& $[s',B_a]$&$\seq$&$ \partial_a c'\ \
,$\\
$[s',\aaux]$&$\seq$&$0\ \ ,$& $[s',\baux]$&$\seq$&$ 0\ \ ,$\\
$\{s',c\}$&$\seq$&$\baux\ \ ,$& $\{s',b\}$&$\seq$&$ \aaux\ \ ,$\\
$\{s',c'\}$&$\seq$&$ 0\ \ ,$& $\{s',b'\}$&$\seq$&$0\ \ .$\\
\end{tabular}
\end{center}
\vskip -0.8truecm
\be\label{EQNBRST}
\ee
Similarly, the transformations generated by the real, vector
super-charges, $s_a$ and $s_a'$ defined by $Q_a\equiv s_a+is_a'$ are found from
(\ref{EQNTRANS})  to be
\begin{center}
\begin{tabular}{rcllcl}
$[s_a,A_b]$&$\seq$&$-\epsilon_{abc} \partial^c c\ \
,$&$[s_a,B_b]$&$\seq$&$-\epsilon_{abc}\partial^c b\ \ ,$\\
$[s_a,\aaux]$&$\seq$&$- \partial_a b\ \ ,$&$[s_a,\baux]$&$\seq$&$-\partial_a c\
\ ,$\\
$\{s_a,c\}$&$\seq$&$0\ \ ,$&$\{s_a,b\}$&$\seq$&$0\ \  ,$\\
$\{s_a,c'\}$&$\seq$&$-B_a\ \ ,$&$\{s_a,b'\}$&$\seq$&$-A_a\ \  ,$\\
$$&$$\\
$[s_a',A_b]$&$\seq$&$-\epsilon_{abc} \partial^c b'\ \
,$&$[s_a',B_b]$&$\seq$&$-\epsilon_{abc}\partial^c c'\ \ ,$\\
$[s_a',\aaux]$&$\seq$&$- \partial_a c'\ \ ,$&$[s_a',\baux]$&$\seq$&$-\partial_a
b'\ \
,$\\
$\{s_a',c\}$&$\seq$&$A_a\ \ ,$&$\{s_a',b\}$&$\seq$&$B_a\ \  ,$\\
$\{s_a',c'\}$&$\seq$&$0\ \ ,$&$\{s_a',b'\}$&$\seq$&$0\ \  .$\\
\end{tabular}
\end{center}
\vskip -0.8truecm
\be\label{EQNSVEC}
\ee
The vector super--charges along with the scalar  BRST and anti-BRST
super--charges satisfy the superalgebra
\be
\{s_a',s_b\}\seq \epsilon_{abc}\partial^c\ \ ,\qquad
\{s_a,s'\}\seq \partial_a\ \ ,\qquad
\{s_a',s\}\seq -\partial_a\ \ ,\label{EQNTBRST}
\ee
with all other combinations vanishing.
The BRST symmetry and the symmetry generated by the vector super--charge,
$s_a'$,
are in agreement with the results of \cite{Luc}.  The transformations of the
anti-BRST and $s_a$ charges were not previously given for the case of $BF$
theories.  Our results verify the
general statement that $s'$ and $s_a'$  may be obtained from $s$ and $s_a$,
respectively,  via
interchanges of ghosts and anti-ghosts.  Due to the first order nature of the
classical $BF$ action this takes the form $c\to b'$, $b'\to -c$, $b\to c'$ and
$c'\to -b$.

Our superalgebras close on-shell only.  Superfield formulations of the
supersymmetric theories in section \ref{ACTIONS} are expected to yield, upon
{\bf{\sl g}}--twisting, off-shell closure of the algebras (\ref{EQNTALG}) and
(\ref{EQNTBRST}).

\vskip 0.5truein\setcounter{equation}{0}
\section{Relating $BF$ to TQFT's}\label{CTQFT}
\forcepar
As mentioned before, twisting a supersymmetric action to  a TQFT requires only
the first step in our {{\bf{\sl g}}}--twisting process in that the Grassmann
parity of the
fields is not changed.
Performing the Grassmann parity change twice is equivalent to the identity.
Thus if we perform a Grassmann parity change on the $BF$ action, we
expect to find a TQFT.  Let us see this explicitly.

Upon making the replacments,
\bea
A_a&\ \rightarrow\ &\rho_{a1}\ \ ,\qquad B_a\ \rightarrow\ \rho_{a2} \ \ ,\nen
\aaux &\ \rightarrow\ & \xi_1\ \ ,\qquad \baux \ \rightarrow\  \xi_2\ \ ,\nen
c&\ \rightarrow\ &\varpi_1\ \ ,\qquad b\ \rightarrow\ \varpi_2\ \ ,\nen
c'&\ \rightarrow\ & \varphi_1\ \ ,\qquad b'\ \rightarrow\  \varphi_2\ \
,\label{EQNTWDW}
\eea
with the Grassmann parity assignments,
\begin{center}
\begin{tabular}{||c|c|c||}\hline
{FIELD}	&SPIN&GRASSMANN PARITY\\ \hline\hline
$\rho_{a i}$	&$1$&odd\\ \hline
$\xi_i$	&$0$&odd\\ \hline
$\varphi_i$	&$0$&even\\ \hline
$\varpi_i$	&$0$&even\\ \hline
\end{tabular}
\end{center}
in the  three dimensional $BF$ action  (\ref{EQNBFT}), we obtain
\be
S'_{TQFT}\seq S_{TQFT}   \mi i \half \int d^3x \epsilon^{abc}\rho_{a
i}\partial_b\rho_{c j} \epsilon^{ij}\ \ ,\label{EQNSFADW}
\ee
where
\be
S_{TQFT}\seq -\int d^3x
\sqrt{g}\sum_{i=1}^2\big[\nabla^a\varpi_i\nabla_a\varphi_i\pls
\rho_i{}^a\nabla_a\xi_i\big]
\ \ .
\ee
Making the same replacements in (\ref{EQNBRST}) yields the BRST transformations
under which $S_{TQFT}'$ is invariant.  We record them for completeness:
\bea
\{s,\rho_{ai}\}&\seq& \partial_a \varpi_{i}\ \ ,\nen
[s,\varpi_i]&\seq&0\ \ ,\nen
[s,\varphi_i]&\seq& -\xi_i\ \ ,\nen
\{s, \xi_i\}&\seq&0\ \ .\label{EQNSTQFT}
\eea
It is then easy to see that
\be
S_{TQFT}\seq\{s,-\int d^3x
\sqrt{g}\sum_{i=1}^2\rho^{a}_{i}\nabla_a\varphi_i\}
\ \ .
\ee
Since the last term in $S_{TQFT}'$ is metric independent, the energy-momentum
tensor from the latter action is $s$--exact.
Of course, starting with this TQFT action and inverting the replacements
(\ref{EQNTWDW})  leads us back to the $BF$ theory.

Alternatively, we could start with our action (\ref{EQNSFN}) and perform the
usual TQFT twist defined to be the map
\be
{{\cal T}}_{TQFT}:\Psi_{\alpha B}\ \rightarrow\ \Psi_{\alpha \beta} \equiv
\inv{\sqrt{2}} \big[\ggmat{^a}_{\alpha\beta} (\rho_{a1}+ i\rho_{a2}) \pls
iC_{\alpha\beta} (\xi_1+i\xi_2)\big]\ \ ,\label{EQNTWDWA}
\ee
which leaves the spin-$0$ fields, $\phi\equiv \inv{\sqrt{2}}(\varphi_1+
i\varphi_2)$
and
$\lambda\equiv \inv{\sqrt{2}}(\varpi_1+ i\varpi_2)$ along with the Grassmann
parity of the
fields unchanged.  With this prescription,  we find that the action,
$S'{}^{\susy}$ becomes $S'_{TQFT}$ up to surface terms.  If we denote
the operation of changing the Grassmann parity of the fields by {\bf {\sl  g}},
then
this information may be encoded in the following diagram:
\begin{center}
\begin{picture}(250,330)
\thicklines
\put(120,300){\vector(-1,-2){39.5}}
\put(130,300){\vector(1,-2){39.5}}
\put(100,210){\vector(1,0){49}}
\put(149,210){\vector(-1,0){49}}
\put(104,300){\framebox(40,20){$S'{}^{\susy}$}}
\put(60,200){\framebox(40,20){$S_{BF}$}}
\put(149,200){\framebox(40,20){$S'_{TQFT}$}}
\put(80,260){${\cal T}_{\sl g}$}
\put(158,260){${\cal T}_{TQFT}$}
\put(122,195){{\sl g}}
\end{picture}
\end{center}
\vskip -2.7truein
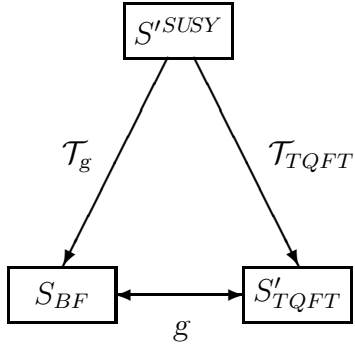
\centerline{Figure 1: The TFT Triangle.}
\vskip 0.75truecm

The last term in $S'_{TQFT}$ does not normally appear in topological sigma
models (even flat ones).  Its presence is idiosyncratic to three dimensions.
It is  invariant under the BRST transformations of eqn. (\ref{EQNSTQFT}).
Although this part of the action has ghost number $-2$, the
full action remains invariant under the $U(1)$ transformation with weights
$(-)^{i}$
for $\rho_{ai}$ and $(-)^{i+1}$ for $\xi_i$.

A similar procedure may be performed using the Chern-Simons action, $S_{CS}^2$,
(\ref{EQNSCSA}).  We find only $S_{TQFT}$ instead of $S_{TQFT}'$; that
is {\bf {\sl g}}:$S_{CS}\to
S_{TQFT}$.  The map is not
invertible as we cannot obtain the Chern-Simons action from
{\bf {\sl g}}: $S_{TQFT}$.  In otherwords, only the gauge fixing and ghost
actions of the Chern-Simons theory may be obtained from $S_{TQFT}$ (or
$S_{TQFT}'$).

\vskip 0.5truein\setcounter{equation}{0}
\section{Conclusions}\label{CONC}
\forcepar

We have defined supersymmetric actions for matter fields which when {\bf{\sl
g}}--twisted (a twist plus Grassmann parity change) yield gauge fixed, abelian
$BF$ theories in three and two dimensions.  In three dimensions, our theory is
$N=4$ supersymmetric while in two dimensions it is $N=2$ supersymmetric.
It has  also been shown how to obtain the gauge fixed Chern-Simons theory via a
\gtwist.  Furthermore, a Donaldson-Witten TQFT is  obtained via the usual
twisting applied to our supersymmetric action.  This yields a scheme for
mapping  the
$BF$ theories into  TQFT's.  For the examples studied we can associate a
topological field theory triangle explicitly illustrating the maps which relate
the supersymmetric, $BF$ and $TQFT$ actions.

The non-abelian case has not been addressed in this work.  It would also be
interesting to check for possible connections between the observables of the
$BF$ theories (linking numbers) and those of the TQFT's.  Indeed, we expect
that our procedure may be generalized to arbitrary dimensional manifolds
(without torsion).
\vskip 0.5truein
\noindent{\Large {\bf Appendix: Conventions}}
\medskip

Our conventions are as follows.
A Majorana spinor, $\psi^\alpha$,  in three dimensions is real and has two
components. Our  gamma matrix conventions in Minkowski space are
$\gamma^a\equiv (\sigma^2,-i\sigma^1,i\sigma^3)$. We have the useful identity
$(\gamma^a\gamma^b)_{\alpha\beta}=\eta^{ab}C_{\alpha\beta} -
i\epsilon^{abc}(\gamma_c)_{\alpha\beta}$.
The charge conjugation matrix,
$C_{\alpha\beta}=\gamma^0=\sigma^2$  acts as
$\psi^\alpha=C^{\alpha\beta}\psi_\beta$ with
$C_{\alpha\beta}C^{\gamma\delta}=\delta_{\alpha}{}^{[\gamma}
\delta_{\beta}{}^{\delta]}$.  Note that since $C$ is imaginary, $\psi_\alpha$
is imaginary.
The metric in Minkowski
space is $\eta={\rm diag}(1,-1,-1)$.
For manifolds with Euclidean signature,  the gamma matrices are
$\gamma^a=(\sigma^2,\sigma^1,\sigma^3)$.  With these conventions,
$\psi^\alpha$($\psi_\alpha$) is still real (imaginary).
The space-time Levi-Cevita tensor is defined by
$\epsilon^{012}\equiv 1$ such that
$\epsilon_{abc}\epsilon^{def}=\delta_a{}^{[d}\delta_b{}^e\delta_c{}^{f]}$.
Internal or $SU(2)$ doublet indices are lowered with the real
sympletic metric $\epsilon_{AB}$ as $\psi^A\epsilon_{AB}=\psi_B$ and raised as
$\epsilon^{AB}\psi_B=\psi^A$.  A bar is used to indicate complex conjugation.

In two dimensions,  our gamma matrices are $\gamma^a=(\sigma^2, -i\sigma^1)$
and $\gamma^3=\sigma^3$.  These satisfy
$\gamma^a\gamma^b=\eta^{ab}-\epsilon^{ab}\gamma^3$ and
$\gamma^3\gamma^a=-\epsilon^{ab}\gamma_b$.  Otherwise, our conventions are
in analogy with three dimensions.

\end{document}